\documentclass[11pt, a4paper, logo]{lime_style}

\usepackage[numbers,sort&compress]{natbib}
\usepackage{xspace}
\usepackage{adjustbox}
\usepackage{array}
\usepackage{float}
\usepackage{placeins}
\usepackage{enumitem}
\usepackage{setspace}
\usepackage{multirow}
\usepackage{makecell}
\usepackage{longtable}
\usepackage{threeparttable}
\usepackage{mathtools}
\usepackage{bm}
\usepackage{dsfont}
\usepackage{subcaption}
\usepackage{tikz}
\usepackage{algorithm}
\usepackage{algpseudocode}
\usepackage{listings}
\usepackage[capitalize,noabbrev]{cleveref}
\usepackage{fontawesome5}
\usepackage[most]{tcolorbox}
\usepackage{caption}
\newcommand{\sysname}{\textsc{CALICO}\xspace}

\newif\ifUseAbbrev
\UseAbbrevfalse

\newcommand{\githublink}[1]{\href{#1}{\faGithub}}
\newcommand{\hflink}[1]{\href{#1}{\raisebox{-0.2ex}{\includegraphics[height=0.95em]{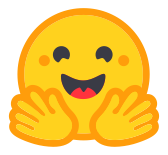}}}}

\usepackage{soul}
\newtcbox{\hlgraytab}{on line, rounded corners, box align=base,colframe=white,size=fbox,arc=2pt, before upper=\strut, top=-2pt, bottom=-4pt, left=-2pt, right=-2pt, boxrule=0pt}

\usepackage{etoolbox}
\newtoggle{release}
\togglefalse{release}

\usepackage{xargs}
\usepackage[colorinlistoftodos,prependcaption,textsize=tiny]{todonotes} 
\newcommandx{\jz}[2][1=]{\todo[linecolor=magenta,backgroundcolor=magenta!25,bordercolor=magenta,#1]{Jieyu: #2}}
\newcommandx{\important}[2][1=]{\todo[linecolor=red,backgroundcolor=red!45,bordercolor=red,#1]{!!! #2}}
\newcommandx{\change}[2][1=]{\todo[linecolor=blue,backgroundcolor=blue!25,bordercolor=blue,#1]{#2}}
\newcommandx{\info}[2][1=]{\todo[linecolor=OliveGreen,backgroundcolor=OliveGreen!25,bordercolor=OliveGreen,#1]{#2}}
\newcommandx{\improve}[2][1=]{\todo[linecolor=Plum,backgroundcolor=Plum!25,bordercolor=Plum,#1]{#2}}
\newcommandx{\add}[2][1=]{\todo[linecolor=Brown,backgroundcolor=Brown!25,bordercolor=Brown,#1]{#2}}
\newcommandx{\later}[2][1=]{\todo[linecolor=Magenta,backgroundcolor=Magenta!25,bordercolor=Magenta,#1]{#2}}
\newcommandx{\here}[2][1=]{\todo[linecolor=Magenta,backgroundcolor=Magenta!25,bordercolor=Magenta,#1]{#2}}
\newcommandx{\solved}[2][1=]{\todo[disable,#1]{#2}}

\def\1{\bm{1}}

\DeclareMathAlphabet{\mathsfit}{\encodingdefault}{\sfdefault}{m}{sl}
\SetMathAlphabet{\mathsfit}{bold}{\encodingdefault}{\sfdefault}{bx}{n}

\title{\centering \sysname: A Human-Centered, Codebook-Aligned System for Annotation}
\renewcommand{\runningtitle}{CALICO: A Human-Centered, Codebook-Aligned System for Annotation}
\date{}

\author[*]{
  Boqin Yuan$^{1,*}$, Xiaoyi Gu$^{*}$, Fiona Li$^{2}$, Chang Wan$^{2}$, Angel Hsing-Chi Hwang$^{2}$, Jieyu Zhao$^{2}$\\
  {\small $^{1}$UC San Diego~~~$^{2}$University of Southern California~~~$^{*}$Equal contribution}\\
  {\large \githublink{https://github.com/calico-annotation/calico-annotation}}~~%
  \vspace{-6pt}
}

\reportnumber{}

\begin{abstract}
Large language models are increasingly used to scale codebook-based annotation in scientific research, but existing workflows provide limited support for translating domain experts' codebooks into reliable, revisable, and auditable prompts. Prompts are often treated as fixed instructions and hidden from annotators, making it difficult for non-technical domain experts to diagnose and correct model behavior when outputs violate codebook guidelines. In this paper, we present \sysname, a human-centered, codebook-aligned annotation workflow that treats prompts as editable, versioned, and optimizable artifacts. \sysname integrates codebook parsing, prompt generation, result inspection, prompt versioning, natural language human feedback, and label-supervised prompt optimization through existing optimizers such as GEPA, MIPROv2, and OPRO, together with our reflection-based optimizer, \textsc{ReflectAgent}. Empirically, we evaluate \sysname on domain-specific AI-companion chatbot conversation codebooks. Across evaluated dimensions, \sysname improves mean held-out performance by $+13.0$ and $+7.4$ absolute points for two coders, respectively. A coder-specificity analysis further suggests that optimized prompts capture coder-specific interpretations rather than only generic codebook clarification. \sysname runs as a web application that takes users from raw codebook materials to inspectable, exportable labels; the website, codebase, and live demo are released at \url{https://calico-annotation.github.io/} under the Apache 2.0 License.

\end{abstract}

\begin{document}

\maketitle

\begin{figure*}[t]
  \centering
  \includegraphics[width=1.0\textwidth]{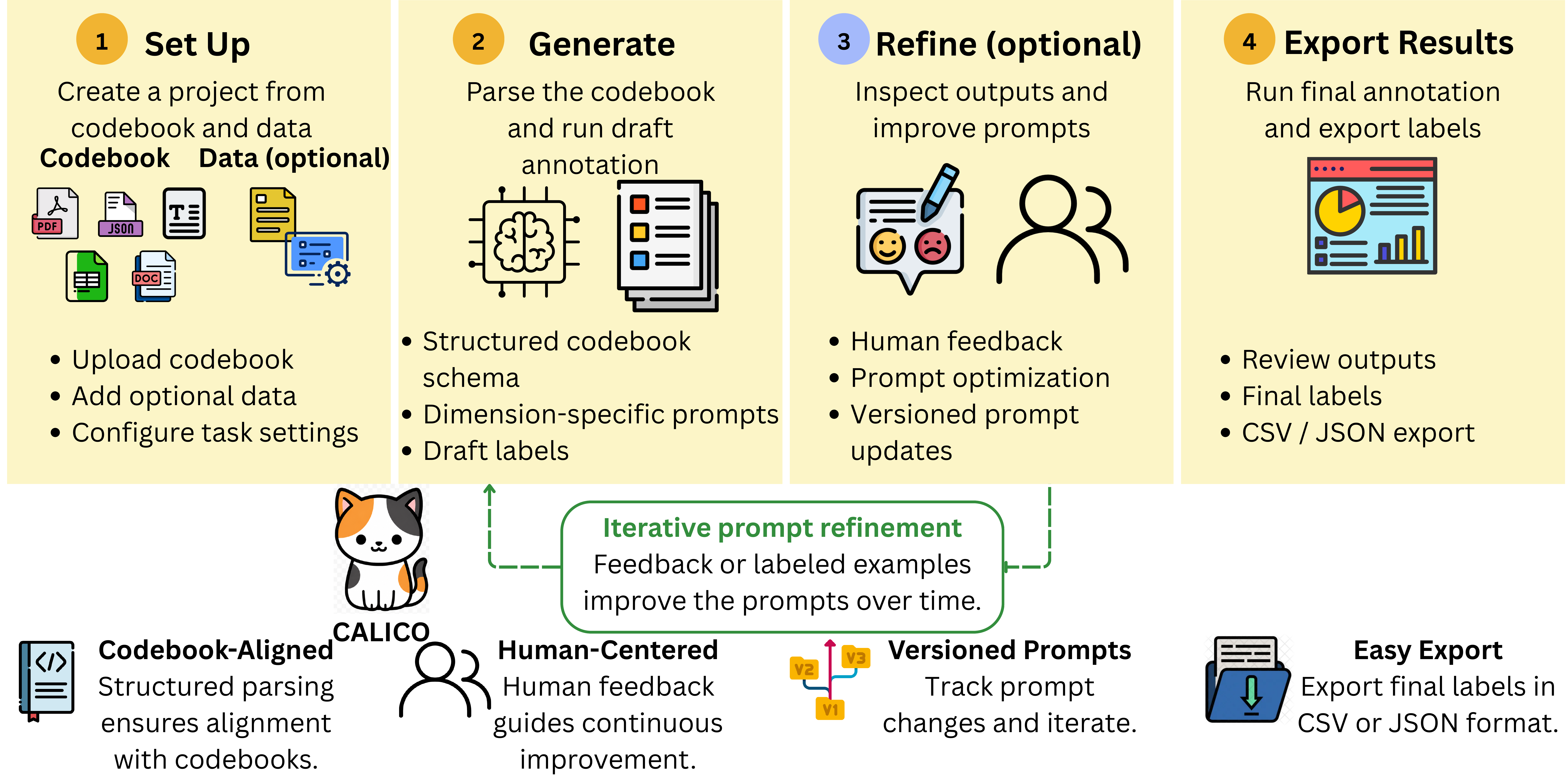}
  \caption{The \sysname workflow. Users create a project from a codebook and optional data, generate dimension-specific prompts, run annotation, inspect outputs, refine prompts when needed, and export final labels. Prompt versions are managed within the project so users can iterate before final annotation.}
  \label{fig:workflow}
  \vspace{-1em}
\end{figure*}

\section{Introduction}
Modern hardware and software systems have made massive amounts of data logs increasingly accessible, providing valuable resources for scientific research in psychology \cite{zhang2025rise}, medicine \cite{abbas2025revolutionizing}, biology \cite{forootani2025bio}, and other domains. In many qualitative and computational social science studies, experts first create a codebook, which is an annotation guide with dimensions, labels, and decision rules, before coding raw data for analysis. Such expert annotation is critical for producing accurate and reliable variables, but it requires substantial time and effort and does not scale well to growing data volumes.

LLMs have made annotation easier to scale through LLM-as-a-judge workflows \cite{zheng2023judging}. In a typical setup, domain experts define an annotation prompt, provide unlabeled data, and ask the LLM to generate labels in batches. However, this convenience does not make codebook-based annotation reliable by default. Annotation quality remains highly contingent on the dataset, task, and how the codebook definitions are operationalized in the prompt \cite{pangakis2023automated, halterman2026codebook}. This is especially challenging for non-technical
domain experts, who may understand the codebook deeply but lack the tools to inspect, debug, and iteratively improve LLM prompts. Existing annotation tools such as Label Studio \citep{labelstudio}, Prodigy \citep{montani2018prodigy}, Humanloop \citep{humanloop}, Doccano \citep{doccano}, and LabelLLM \citep{labelllm_software} can use LLMs to pre-fill labels for experts to accept or edit, but the pre-fill prompt typically does not adapt to the corrections it receives.

To address these challenges, we present \sysname (\underline{C}odebook-\underline{A}ligned \underline{L}LM-assisted \underline{I}terative  \underline{C}oding and \underline{O}ptimization), a human-centered, codebook-aligned workflow for LLM-assisted annotation. \sysname is designed for domain experts who have data to annotate and a codebook to apply, without requiring them to manually engineer and manage prompts across annotation rounds. \sysname supports prompt improvement under two common conditions. \textbf{1) Cold-start annotation:} No expert annotations are available initially. \sysname generates the prompt based on the codebook, produces first-pass annotations, and updates its prompts as experts refine those outputs using human natural language feedback. \textbf{2) Label-supervised optimization:} Existing expert labels are available. They are used to guide configurable optimizers including GEPA \cite{gepa}, MIPROv2 \cite{miprov2}, and OPRO \cite{opro}, together with our reflection-based optimizer, \textsc{ReflectAgent} to improve the prompts. In this setting, we also provide evaluation metrics to help users assess each prompt update. Human feedback remains available as a complementary correction signal. With a human-centered focus, \sysname treats codebooks, prompts, annotation runs, feedback, and prompt versions as auditable artifacts throughout the workflow, allowing experts to inspect changes, manage prompt history, and reuse improved prompts in later sessions.

In summary, we propose \sysname, a human-centered, codebook-aligned workflow. Through empirical experiments, we demonstrate that \sysname improves mean held-out performance by $+13.0$ and $+7.4$ points for two coders respectively.

\section{Related Work}
\label{sec:related}

\paragraph{Human-LLM collaborative annotation.}

LLMs are increasingly used as scalable annotators \citep{zheng2023judging, tan2024large, shankar2024validators}. Researchers have explored LLM-assisted qualitative coding using expert defined codebooks \citep{dunivin2025scaling, tai2024exam, xiao2023supporting, than2025updating}. LLM-assisted annotation has also been applied to domains such as psychology \cite{zhang2025rise} and media analysis \cite{horych2025promises}. Annotation platforms like Label Studio, Prodigy \citep{montani2018prodigy}, Doccano, and LabelLLM \citep{labelllm_software} support dataset management and AI-assisted pre-annotation, where model-generated labels can be reviewed or corrected by humans.

Recent systems explore how humans and LLMs can collaborate during annotation. MEGAnno+ \citep{kim2024megannoplus} manages LLM agents, annotation jobs, metadata, and human verification of LLM labels. Lapras \citep{wang2024human} uses LLM labels and explanations, then trains a verifier to select likely incorrect labels for human re-annotation. PDFChatAnnotator \citep{tang2024pdfchatannotator} lets users guide LLM-based extraction from PDF catalogs through examples and iterative requirement adjustment. At the same time, LLM assistance can affect human judgments in subjective codebook tasks, shifting label distributions toward model suggestions \citep{schroeder2025just}. \sysname builds on this human-centered motivation, but differs from workflows centered on verifying or correcting LLM-generated labels by making the prompts that operationalize domain codebooks inspectable, refinable, and versioned.

\paragraph{Agentic workflows and prompt refinement.}
Researchers have explored agentic workflows, codebook refinement, and prompt optimization to further improve LLM annotation performance and better align it with human annotators. Agentic systems like CrowdAgent \citep{crowdagent2025} and EvoAgentX \citep{evoagentx2025} coordinate or evolve multi-agent workflows with reasoning, memory, and actions, while Co-DETECT \citep{codetect2025} uses LLMs to surface edge cases from a sketch codebook and help users refine codebook rules. Separately, prompt optimizers such as DSPy \citep{khattab2023dspycompilingdeclarativelanguage}, OPRO \citep{opro}, MIPROv2 \citep{miprov2}, and GEPA \citep{gepa} improve prompts from labeled examples, but are typically used as standalone optimization methods. \sysname connects these directions in a codebook-aligned annotation workflow. 

\section{System Design}
\label{sec:system_design}

Figure \ref{fig:workflow} illustrates the end-to-end \sysname workflow. \sysname follows a lightweight, codebook-centered design: experts start from the codebook that already defines their annotation task, and the system turns it into editable prompts for LLM-assisted annotation. After inspecting model outputs, experts can refine these prompts through natural language feedback, describing systematic errors, boundary cases, or preferred labeling behavior. When labeled examples are available, users can also run label-supervised optimizers and inspect the resulting prompt updates, artifacts, and evaluation metrics.

To support this, \sysname has four modular components: Codebook Parsing Agent (\S~\ref{sec:codebook_parsing_agent}), Prompt Generator (\S~\ref{sec:prompt_generator}), Human-in-the-Loop Feedback (\S~\ref{sec:human_in_the_loop}), and Prompt Optimizer (\S~\ref{sec:prompt_optimizer}).

\subsection{Codebook Parsing Agent}
\label{sec:codebook_parsing_agent}

The Codebook Parsing Agent is designed to make project setup lightweight for domain experts. Instead of requiring users to clean or convert their materials into a fixed template, \sysname accepts codebooks and annotation data in common document, spreadsheet, and structured formats, including PDF, DOCX, XLSX, CSV, JSON, and plain text. It applies format-specific preprocessing to preserve useful structure, such as spreadsheet layouts and existing label columns, while
extracting candidate coding dimensions, labels, definitions, instructions, examples, and labeling modes.

After preprocessing, the agent consolidates these elements into a canonical codebook schema used by the rest of the workflow. This schema also enables automatic validation and alignment: \sysname flags missing or inconsistent codebook fields, detects structural issues such as duplicated or gated labels, and aligns uploaded label values to the closest valid schema entries. Users can review and edit the parsed schema and validation warnings before activating it for following annotations.

\subsection{Prompt Generator}
\label{sec:prompt_generator}

Once a codebook is accepted, the Prompt Generator turns its schema into executable prompts for LLM-assisted annotation, reducing the need for users to manually translate codebook rules into prompt instructions. For each coding dimension, \sysname assembles the relevant labels, definitions, task instructions, examples, and output constraints into an initial prompt. These prompts remain visible and editable in the interface, so experts can inspect how the codebook has been operationalized before running annotation.

The generator also carries codebook constraints into the annotation prompt and runtime configuration. For example, it adds explicit guidance for ``No label'' options when they appear in the codebook, and records hierarchical or gated label restrictions so later predictions respect dependencies among coding dimensions. This design keeps the annotation process aligned with the original codebook while presenting users with a simple editable prompt for each dimension.

\subsection{Human-in-the-Loop Feedback}
\label{sec:human_in_the_loop}

Cold-start annotation is common in domain-specific settings, where experts often have a codebook and unlabeled data but few or no existing annotations. \sysname addresses this setting through a human-in-the-loop feedback workflow, using model-generated annotations as drafts for expert review rather than final labels. Users inspect annotated examples and provide free-text feedback about systematic errors, boundary cases, or preferred labeling behavior (as shown in Figure~\ref{fig:human_feedback_main_paper} and more details in Appendix~\ref{sec:appendix_interface}).

\begin{figure*}[t]
    \centering
    \includegraphics[width=0.8\textwidth]{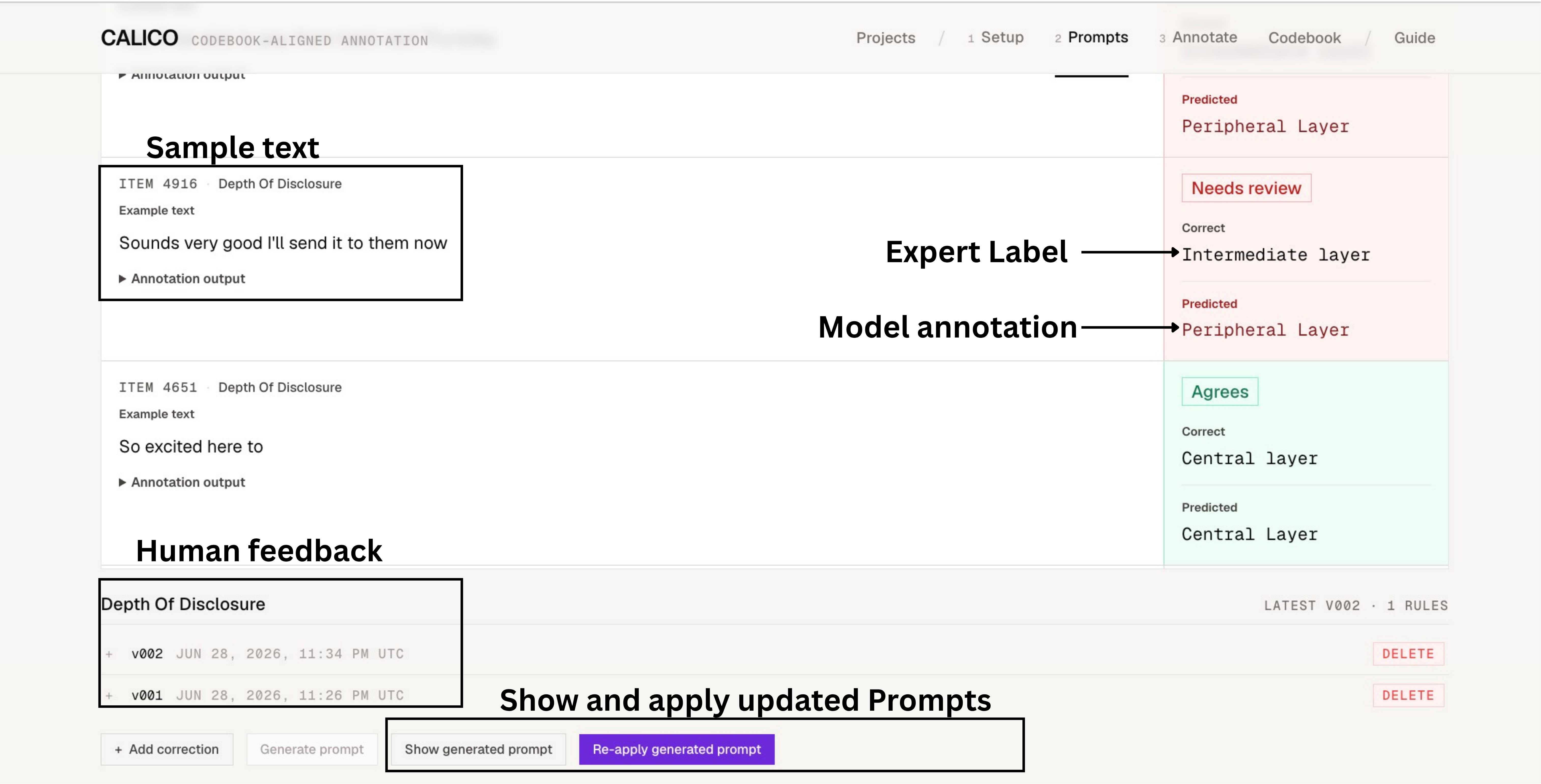}
    \caption{Feedback and prompt-update interface. Users provide natural language feedback and review the proposed prompt revision before applying it as a new prompt version.}
    \label{fig:human_feedback_main_paper}
    \vspace{-1em}
\end{figure*}

Internally, \sysname does not apply free-text feedback directly to the prompt. Instead, it summarizes feedback into reusable calibration guidance, such as label-boundary rules or corrections to common failure patterns. The guidance is written to memory tied to the current project, codebook version, and coding dimension, allowing feedback to accumulate within relevant annotation context. Before updating the prompt, \sysname shows the proposed revision and its diff from the current prompt, and only approved changes are saved as a new prompt version.

When reference labels are available, the interface can also show them alongside model predictions, and the same feedback workflow can be used as a complementary correction signal in label-supervised projects.

\subsection{Prompt Optimizer}
\label{sec:prompt_optimizer}

\paragraph{Optimizers.}
\sysname supports label-supervised prompt optimization after prompt generation. This means that when users have a small set of expert labeled examples for a coding dimension, \sysname can use those labels to improve the prompt. Given one prompt, the labels allowed by the codebook, and the labeled examples, each optimizer returns a new candidate prompt and its optimization results. The current implementation includes GEPA~\citep{gepa}, MIPROv2~\citep{miprov2}, OPRO~\citep{opro}, and our reflection-based optimizer, \textsc{ReflectAgent}. MIPROv2 searches over instructions and few-shot examples, GEPA performs DSPy-based reflective prompt optimization, and OPRO asks the LLM to propose new prompts from prior results. Because all optimizers use the same input format and execution pipeline, \sysname lets users compare optimization backends within one project.

\paragraph{ReflectAgent.}
\textsc{ReflectAgent} aims to improve prompt accuracy for one coding dimension over several rounds. In each round, the model uses the current prompt to label the training examples. A pattern-extraction module then reads the errors and abstracts candidate calibration rules. Unlike human feedback, which starts from user-written corrections, \textsc{ReflectAgent} uses existing expert-labeled examples to proactively mine errors and propose prompt rules. Each rule captures a label boundary, target labels, optional cue families, and an annotator-readable instruction. The extractor is constrained to generalize from failures rather than copy examples verbatim, so the resulting rules capture reusable decision principles instead of surface-specific demonstrations. \textsc{ReflectAgent} folds accepted rules into the candidate prompt, producing an optimized prompt that remains editable and inspectable by the user.

\paragraph{Evaluation and rollback.}
For label-supervised prompt optimization, \sysname splits labeled examples into deterministic stratified train, validation, and held-out test sets. Training examples drive prompt search or failure mining, while validation performance is used to select candidate updates and trigger rollback when performance regresses beyond a tolerance. After optimization completes, \sysname evaluates the initial and optimized prompts on the held-out test set exactly once. For single-label dimensions, \sysname reports accuracy, macro-$F_1$, and per-class metrics; for multi-label dimensions, it reports micro-$F_1$ and per-label metrics. The system also runs a leakage audit that checks whether validation or test examples appear verbatim in the final prompt. This separation lets users inspect optimization trajectories while preserving a held-out estimate of generalization.

\begin{table*}[!t]
  \centering
  \small
  \setlength{\tabcolsep}{4pt}
  \newcommand{\ms}[2]{$#1_{\,\textcolor{gray}{\pm#2}}$}
  \newcommand{\bms}[2]{$\mathbf{#1}_{\,\textcolor{gray}{\pm#2}}$}
  \newcommand{\pos}[1]{\textcolor{green!50!black}{\ensuremath{#1}}}
  \newcommand{\loss}[1]{\textcolor{red!70!black}{\ensuremath{#1}}}
  \begin{tabular*}{\textwidth}{@{\extracolsep{\fill}}lcccccc@{}}
    \toprule
     & \multicolumn{3}{c}{\textbf{Coder A}} & \multicolumn{3}{c}{\textbf{Coder B}} \\
    \cmidrule(lr){2-4}\cmidrule(lr){5-7}
    \textbf{Dimension} & Zero-shot & + ReflectAgent & $\Delta$ & Zero-shot & + ReflectAgent & $\Delta$ \\
    \midrule
    \multicolumn{7}{@{}l}{\textbf{Self-Disclosure}\quad\textit{(single-label, accuracy \%)}}\\
    \quad Level of disclosure
      & \ms{68.1}{2.7} & \bms{70.0}{1.7} & \pos{+1.9}
      & \ms{61.7}{0.6} & \bms{76.6}{5.1} & \pos{+14.9} \\
    \quad Disclosure as confession
      & \ms{52.6}{1.2} & \bms{82.0}{8.7} & \pos{+29.4}
      & \ms{49.6}{3.8} & \bms{50.1}{1.9} & \pos{+0.5} \\
    \quad Depth of disclosure
      & \bms{61.0}{6.6} & \ms{58.7}{2.4} & \loss{-2.3}
      & \ms{58.2}{2.5} & \bms{61.2}{7.2} & \pos{+3.0} \\
    \quad Intimacy of self-disclosure
      & \ms{52.2}{3.2} & \bms{75.5}{3.1} & \pos{+23.3}
      & \ms{54.9}{2.5} & \bms{66.0}{3.5} & \pos{+11.1} \\
    \quad\textit{mean}
      & \textit{58.5} & \textbf{\textit{71.5}} & \pos{+13.0}
      & \textit{56.1} & \textbf{\textit{63.5}} & \pos{+7.4} \\
    \addlinespace
    \multicolumn{7}{@{}l}{\textbf{AI Behavior}\quad\textit{(multi-label, micro-$F_1$ \%)}}\\
    \quad Listening strategy
      & \ms{71.6}{4.5} & \bms{73.2}{4.0} & \pos{+1.6}
      & \ms{61.3}{3.4} & \bms{68.8}{4.7} & \pos{+7.5} \\
    \bottomrule
    \vspace{-1em}
  \end{tabular*}

  \caption{Per-coder agreement with each coder's held-out labels, comparing the zero-shot prompt with \textsc{ReflectAgent}. Scores are accuracy (\%) for Self-Disclosure and micro-$F_1$ (\%) for AI Behavior. Results are reported as mean$\pm$standard deviation over three seeds. Bold indicates the better score between zero-shot and \textsc{ReflectAgent}; green and red $\Delta$ values indicate absolute gains and drops from zero-shot. For AI Behavior, we report results on the Listening strategy theme.}

  \label{tab:main}
\end{table*}

\section{Experiments}

We run \sysname on AI-companion chatbot analysis with two expert codebooks as a case study. Our goal is to evaluate whether \sysname's label-supervised prompt refinement can improve over a common LLM-assisted annotation baseline: a zero-shot LLM annotator initialized from the codebook. We focus on two questions:

\textbf{RQ1: Does \sysname improve held-out annotation performance over a zero-shot (vanilla) LLM annotator initialized from the codebook?}

\textbf{RQ2: Are the gains specific to the coder whose labels were used, rather than only reflecting generic codebook clarification?}

\subsection{Experiment Settings}

\paragraph{Codebooks.} We evaluate \sysname on two codebooks from an AI-companion conversation analysis project. The \emph{Self-Disclosure} codebook is used to annotate user messages along four single-label dimensions: level \citep{balani2015detecting, zhang2025rise}, depth \citep{skjuve2023longitudinal}, intimacy, and confession \citep{croes2024digital}. The \emph{AI Behavior} codebook is applied to annotate AI messages under two multi-label themes: listening strategy \citep{bodie2012listening}, with seven labels, and support type \citep{yin2024ai}, with two labels.

\paragraph{Datasets.}
The data were independently labeled by two trained expert coders, i.e., annotators who apply the codebooks, denoted Coder A and Coder B. For Self-Disclosure, Coder A and Coder B labeled 333 and 330 user messages in total, with 124--330 labeled items per single-label dimension. For AI Behavior, they labeled 340 and 335 AI messages in total; for the reported Listening strategy theme, Coder A and Coder B labeled 291 and 231 items, respectively. Notably, we observed that their annotations show substantial disagreement: for the \emph{Self-Disclosure} codebook, raw agreement is $67.4\%$ on Level and only $34.8\%$ on Intimacy. Therefore, rather than adjudicating these labels into a single gold dataset, we evaluate each coder separately. For each coder and dimension, that coder's labels define the reference labels, and performance is measured as agreement with that coder's held-out labels. In each run, we split the coder-specific labeled set into train, validation, and test partitions using the deterministic seed described in Appendix~\ref{sec:appendix_implementation}; the test split is held out from optimization and scored only once.

\paragraph{Metrics.}
For single-label dimensions, we report accuracy, defined as the percentage of held-out items where the model prediction matches the target coder's label. For multi-label dimensions, we report micro-$F_1$ over held-out items.

\subsection{Results}

Table~\ref{tab:main} presents held-out agreement with each coder's labels before (zero-shot) and after applying \textsc{ReflectAgent}, using \texttt{gpt-5.4-mini} as the annotation model (macro-$F_1$ numbers in Appendix \ref{sec:appendix_implementation}). We report absolute changes in percentage points (pp). Across the four Self-Disclosure dimensions, \textsc{ReflectAgent} improves mean accuracy by $+13.0$ pp for Coder A and $+7.4$ pp for Coder B. The improvements are not uniform across dimensions. For Coder A, the largest gains appear on Confession ($+29.4$ pp) and Intimacy ($+23.3$ pp), while for Coder B, the largest gains appear on Level ($+14.9$ pp) and Intimacy ($+11.1$ pp). The pattern suggests that the optimizer is not learning a single generic clarification, but is adapting to the labels used for supervision. For the multi-label AI Behavior codebook, \textsc{ReflectAgent} also improves Listening strategy micro-$F_1$, with a small gain for Coder A ($+1.6$ pp) and a larger gain for Coder B ($+7.5$ pp). Overall, this answers \textbf{RQ1}: with \textsc{ReflectAgent}, \sysname improves average held-out agreement for both coders.

\begin{table}[t]
  \centering\footnotesize
  \setlength{\tabcolsep}{2.5pt}
  \newcommand{\ms}[2]{$#1_{\,\textcolor{gray}{\pm#2}}$}
  \newcommand{\bms}[2]{$\mathbf{#1}_{\,\textcolor{gray}{\pm#2}}$}
  \begin{tabular}{@{}lccccc@{}}
    \toprule
    \textbf{Dim.} & \textbf{ZS} & \textbf{OPRO} & \textbf{GEPA} & \textbf{MIPRO} & \textbf{RA} \\
    \midrule
    Level       & 68.1 & \ms{72.4}{4.6} & \bms{74.8}{1.6} & \ms{72.2}{3.0} & \ms{70.0}{1.7} \\
    Confession  & 52.6 & \ms{52.6}{2.6} & \ms{77.8}{5.5} & \bms{85.2}{7.3} & \ms{82.0}{8.7} \\
    Depth       & 61.0 & \ms{58.7}{6.3} & \bms{64.8}{3.4} & \ms{62.0}{5.0} & \ms{58.7}{2.4} \\
    Intimacy    & 52.2 & \ms{62.3}{4.6} & \bms{78.0}{2.4} & \ms{76.1}{3.6} & \ms{75.5}{3.1} \\
    \midrule
    \textit{mean} & \textit{58.5} & \textit{61.5} & \textit{73.8} & \textbf{\textit{73.9}} & \textit{71.5} \\
    \bottomrule
  \end{tabular}
  \caption{Comparison of prompt optimization backends on Coder A's Self-Disclosure dimensions. Scores are accuracy (\%); optimized methods report mean$\pm$standard deviation over three seeds. All methods start from the same zero-shot (ZS) prompt and use identical data splits. Bold indicates the best score in each row. RA denotes \textsc{ReflectAgent}.}
  \label{tab:optimizers}
  \vspace{-1em}
\end{table}

\paragraph{Performance of Optimizers.}
Beyond \textsc{ReflectAgent}, \sysname integrates several existing prompt optimizers, including OPRO, MIPROv2, and GEPA. To evaluate this label-supervised optimization interface, we run all optimizers on Coder A's Self-Disclosure dimensions using the same zero-shot prompt and identical three-seed splits. As shown in Table~\ref{tab:optimizers}, most optimizers improve over the zero-shot prompt. MIPROv2 and GEPA achieve the strongest mean accuracy ($73.9$ and $73.8$), \textsc{ReflectAgent} is comparable ($71.5$), and OPRO gives a smaller improvement ($61.5$). These results show that \sysname can support multiple optimization backends within the same codebook-centered workflow. Unlike the existing optimizers, \textsc{ReflectAgent} produces an inspectable and editable rule library, making the optimization artifact easier for domain experts to review, revise, and reuse across annotation sessions.

\paragraph{Coder-specificity analysis.}
To answer \textbf{RQ2}, we test whether prompt refinement captures coder-specific interpretations rather than only generic codebook clarification. We focus on the Intimacy dimension, where the two coders show substantial disagreement, with only $34.8\%$ raw agreement. Starting from the same zero-shot prompt, we run \textsc{ReflectAgent} separately using each coder's labels, then evaluate each tuned prompt against both coders' held-out labels on the same shared items. If the learned rules only clarified the codebook in a generic way, both tuned prompts should perform similarly against both coders. Instead, Table~\ref{tab:spec} shows a diagonal pattern: the Coder A-tuned prompt achieves $75.2\%$ accuracy against Coder A but only $41.0\%$ against Coder B, while the Coder B-tuned prompt achieves $71.4\%$ against Coder B but only $51.4\%$ against Coder A. This indicates that each prompt performs best when evaluated against the coder whose labels were used for tuning.

\begin{table}[t]
  \centering\small
  \begin{tabular}{@{}lcc@{}}
    \toprule
    \textbf{Tuned on} & \textbf{Eval. vs Coder A} & \textbf{Eval. vs Coder B} \\
    \midrule
    Coder A labels & \textbf{75.2} & 41.0 \\
    Coder B labels & 51.4 & \textbf{71.4} \\
    \bottomrule
  \end{tabular}
    \caption{Coder-specificity analysis on the Intimacy dimension. Values are mean held-out accuracy (\%) over three seeds on the shared held-out set ($n{=}35$). Rows indicate the coder used for tuning, and columns indicate the coder used for evaluation. Matched-coder prompts outperform swapped-coder prompts by $+23.8{\pm}7.5$ pp for Coder A and $+30.5{\pm}10.8$ pp for Coder B, suggesting that \textsc{ReflectAgent} captures coder-specific label boundaries.}
  \label{tab:spec}
  \vspace{-1em}
\end{table}

\section{Conclusion}
We presented \sysname, a human-centered, codebook-aligned workflow for LLM-assisted annotation. \sysname supports codebook parsing, prompt generation, draft annotation, result inspection, prompt refinement, version management, and label export. It provides natural language feedback for cold-start settings and label-supervised optimization when labels are available. Our empirical evaluation on an AI-companion conversation codebook under label-supervised optimization improves annotation performance over a zero-shot prompt for both coders on average, and a coder-specificity analysis suggests that optimized prompts can capture coder-specific label boundaries.

\section*{Limitations}
Our evaluation focuses on offline agreement for one domain-specific codebook and does not include a live user study, so we do not claim cross-domain generalization or measured usability gains. \sysname currently updates prompts and rule libraries but does not fine-tune annotation models; future work will study task-specific fine-tuning for scalability. The preprocessing also offers limited control over annotation units, such as splitting long messages into multiple rows. Finally, \sysname targets discrete codebook labels; soft label distributions require a different output workflow.



\bibliographystyle{unsrtnat}
\bibliography{reference}

\clearpage
\appendix

\section{Implementation Details}
\label{sec:appendix_implementation}

\paragraph{Split and leakage guard.}
For each coder and dimension, we create a deterministic stratified train/validation/test split ($15\%/42\%/43\%$). The small train fraction reflects the low-label regime \sysname targets, while the larger validation and test splits make prompt selection and held-out evaluation more stable. Each split is seeded by hashing \texttt{coder|dimension|seed\_index} with SHA-256 and using the first 8 hex digits as the integer seed; multi-run results use seed indices $0$, $1$, and $2$, while single-seed appendix results use seed index $0$. The optimizer uses only the train and validation sets: training examples drive failure mining or prompt search, while validation examples select updates and trigger rollback when performance drops by more than $\epsilon{=}0.005$. The held-out test set is disjoint by example ID and scored once after optimization. Each \textsc{ReflectAgent} rule contains an identifier, dimension, label boundary, target labels, cues, and an annotator-readable instruction; Figure~\ref{fig:rule} shows an example.

\begin{figure}[h]
\footnotesize
\begin{verbatim}
{
 "id": "level-001",
 "dimension": "Level",
 "boundary": "generic agreement vs.
              minimal personal content",
 "target_labels": ["No", "Low"],
 "positive_cues": [
   "names a personal preference",
   "narrates a first-person event"],
 "negative_cues": [
   "purely generic agreement",
   "third-person commentary"],
 "rule": "If the utterance carries any
   first-person anchor, label Low;
   reserve No for empty agreement."
}
\end{verbatim}
\caption{Example entry in the \textsc{ReflectAgent} rule library. Each rule records a reusable decision boundary that can be inspected and edited by annotators.}
\label{fig:rule}
\end{figure}

\paragraph{Macro-$F_1$ scores.}
Table~\ref{tab:macro} reports macro-$F_1$ for the Self-Disclosure dimensions using one split seed. The trends broadly match the accuracy results in Table~\ref{tab:main}: \textsc{ReflectAgent} improves the largest-gain dimensions for each coder, while changes on the remaining dimensions are smaller.

\begin{table}[t]
  \centering\small
  \begin{tabular}{@{}lcccc@{}}
    \toprule
    & \multicolumn{2}{c}{\textbf{Coder A}} & \multicolumn{2}{c}{\textbf{Coder B}} \\
    \cmidrule(lr){2-3}\cmidrule(lr){4-5}
    Dimension & ZS & +RA & ZS & +RA \\
    \midrule
    Level       & 0.581 & \textbf{0.598} & 0.659 & \textbf{0.858} \\
    Confession  & 0.482 & \textbf{0.686} & 0.457 & \textbf{0.529} \\
    Depth       & \textbf{0.569} & 0.549 & 0.581 & \textbf{0.609} \\
    Intimacy    & 0.525 & \textbf{0.643} & 0.532 & \textbf{0.611} \\
    \bottomrule
  \end{tabular}
  \caption{Macro-$F_1$ on the Self-Disclosure dimensions for one split seed. ZS denotes the zero-shot prompt, and +RA denotes the prompt after \textsc{ReflectAgent}. Bold indicates the better score between ZS and +RA for each coder and dimension.}
  \label{tab:macro}
\end{table}

\section{Robustness across annotation models}
\label{sec:appendix_models}
To check that \textsc{ReflectAgent}'s gains are not tied to one base model, we repeat the Coder A Intimacy experiment across four models, including API vendors and a self-hosted open model, on identical splits. As Table~\ref{tab:models} shows, every model improves by $+20.8$ to $+34.6$ pp, with the three API models converging to $73$--$78\%$ from different zero-shot starts. The consistent gains suggest codebook-specific alignment rather than a single-model artifact.

\begin{table}[h]
  \centering\small
  \begin{tabular}{@{}lccc@{}}
    \toprule
    Model & Zero-shot & +RA & $\Delta$ \\
    \midrule
    \texttt{gpt-5.4-mini} & 52.2 & \textbf{75.5} & +23.3 \\
    \texttt{gpt-5.5} & 45.3 & \textbf{78.3} & +33.0 \\
    \texttt{gemini-3.1-pro} & 38.4 & \textbf{73.0} & +34.6 \\
    \texttt{Qwen3-8B} & 37.1 & \textbf{57.9} & +20.8 \\
    \bottomrule
  \end{tabular}
  \caption{Intimacy agreement with Coder A (accuracy \%) across annotation models, zero-shot versus \textsc{ReflectAgent} (+RA), on identical stratified splits across three seeds. \texttt{Qwen3-8B} is self-hosted via vLLM.}
  \label{tab:models}
\end{table}

\FloatBarrier
\section{Interface Details}
\label{sec:appendix_interface}
\begin{figure*}[t]
    \centering
    \includegraphics[width=\textwidth]{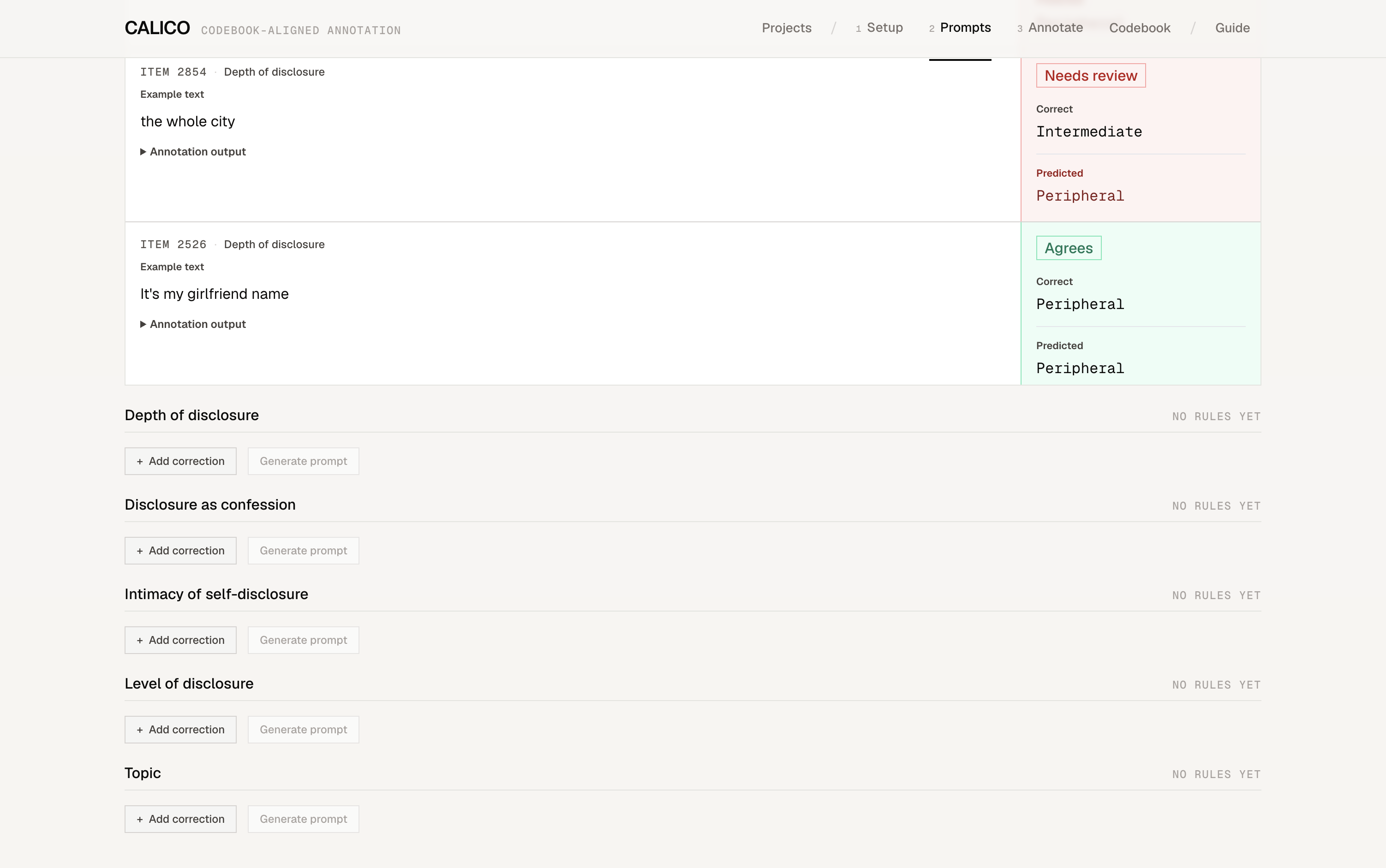}
    \caption{Feedback and prompt-update interface. Users provide natural language feedback and review the proposed prompt revision before applying it as a new prompt version.}
    \label{fig:human_feedback}
\end{figure*}

\paragraph{Human-in-the-Loop Feedback.}
Figure~\ref{fig:human_feedback} shows the feedback workflow in \sysname. Users first inspect model-generated annotations together with the input item and, when available, the reference label. They can then provide natural language feedback about systematic errors or label-boundary decisions. \sysname converts this feedback into structured calibration guidance and shows the proposed prompt update before users apply it.

\begin{figure*}[t]
    \centering
    \includegraphics[width=\textwidth]{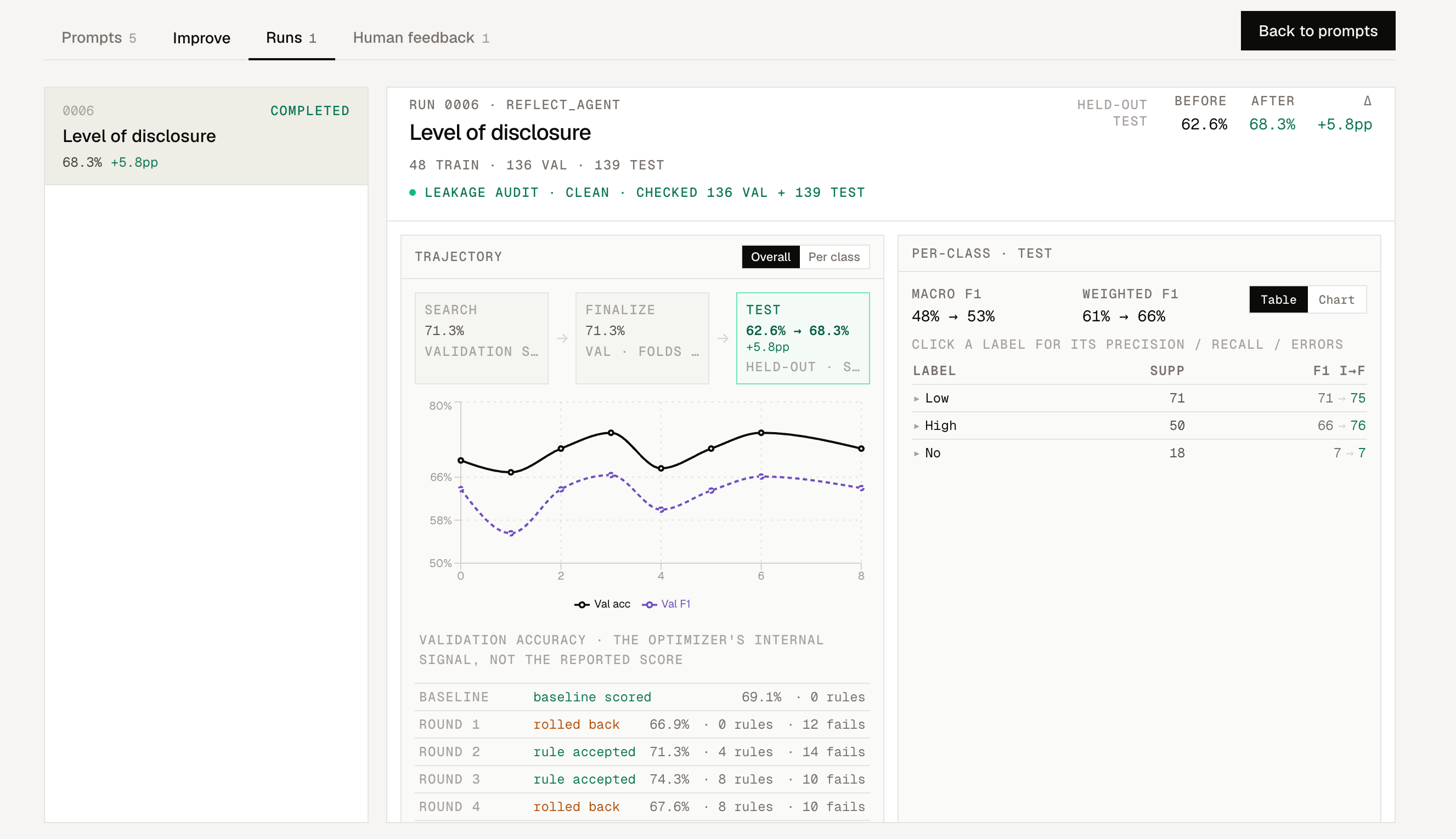}
    \caption{The label-supervised prompt improvement view in \sysname, for a \textsc{ReflectAgent} run on the \emph{Level of disclosure} dimension.}
    \label{fig:improve_run}
\end{figure*}

\paragraph{Prompt Optimization.}
Figure~\ref{fig:improve_run} shows the label-supervised prompt improvement view in \sysname for a \textsc{ReflectAgent} run on the \emph{Level of disclosure} dimension. The header reports the held-out test improvement ($62.6\% \to 68.3\%$, $+5.8$ pp). The trajectory panel plots the per-round validation signal and lists each round's accept-or-rollback decision and rule count. This inspectable view lets domain experts verify what changed and whether it generalizes before adopting a new prompt.

\end{document}